\documentstyle[preprint,aps,eqsecnum]{revtex}
\tighten
\begin{document}
\title{Topology, Locality, and Aharonov-Bohm Effect with Neutrons}
\author{Murray Peshkin}
\address{Physics Division, Argonne National Laboratory, Argonne, IL
60439-4843 USA}
\author{H. J. Lipkin}
\address{Department of Particle Physics, Weizmann Institute of Science,
Rehovot 76100, Israel \\
and \\ School of Physics, Raymond and Beverly Sackler Faculty of Exact
Sciences \\ Tel-Aviv University, Tel Aviv, Israel}

\maketitle
\begin{abstract}
Recent neutron interferometry experiments have been interpreted as
demonstrating a new topological phenomenon similar in principle to the usual
Aharonov-Bohm (AB) effect, but with the neutron's magnetic moment replacing
the electron's charge.  We show that the new phenomenon, called Scalar AB
(SAB) effect, follows from an ordinary local interaction, contrary to the
usual AB effect, and we argue that the SAB effect is not a topological effect
by any useful definition.  We find that SAB actually measures an apparently
novel spin autocorrelation whose operator equations of motion contain the
local torque in the magnetic field.  We note that the same remarks apply to
the Aharonov-Casher effect.
\end{abstract}

\newpage

\section{The Aharonov-Bohm effect with electrons}

In the Aharonov-Bohm effect (AB),\cite{aharonov59,peshkin89} idealized
in Fig. 1, the motion of an electron in a
Mach-Zehnder interferometer is influenced by electromagnetic fields even though
 the electron
experiences no local, contemporaneous Maxwell field.  That comes about because
 the Hamiltonian

\begin{equation}
H = {1 \over 2m} ({\mbox{\boldmath$p$}} + {e \over c} {\mbox{\boldmath$A$}})^2
-
 eV
\end{equation}
contains the gauge fields V and $\bf{A}$, which have nonvanishing values at
some
 points
in the domain
of the electron's position $\bf{r}$.  For AB effect, we can ignore the
 electron's spin.
Then the operator
equations of motion for the only observables are

\begin{equation}
\dot{\mbox{\boldmath$r$}} = {\mbox{\boldmath$v$}} \,\,\,\,\,\,\,\,\,\,\,\ m
\dot{\mbox{\boldmath$v$}} = 0\,,
\end{equation}
those of a free particle, containing no electromagnetic fields.  However, in
 quantum mechanics
the
equations of motion alone do not determine the dynamics. In the magnetic AB
 effect (Fig. 1a),
the partial waves in the two arms of the interferometer acquire a relative
phase  shift
${\delta}{\phi}$ given by

\begin{equation}
\delta \phi = {e \over \hbar c} \oint A \cdot d{\mbox{\boldmath$r$}} = {e\Phi
 \over \hbar c}\,,
\end{equation}
where $\Phi$ is the flux through the solenoid.

In the electric AB effect (Fig. 1b), the two arms of the interferometer carry
 the electron
through conducting cylinders that shield the electron from an electric field.
 While the split
wave
packet is deep within one cylinder or the other, potentials $V_{1}$ and $V_{2}$
 are applied
to the two
cylinders.  That causes a relative phase shift given by

\begin{equation}
\delta \phi = {e \over \hbar} (\Delta V) \tau\,,
\end{equation}
where ${\Delta}V=V_{1}-V_{2}$ and ${\tau}$ is the length of the time interval
 during which
${\Delta}V$ is different from
zero.

In both cases, the relative phase shift is measured by the outbound intensities

\begin{equation}
I_{1} = I{cos^2} (d\phi) \,\,\,\,\,\,\,\ I_2 = I{sin^2}(d\phi)\,.
\end{equation}
The AB effect is nonlocal in that the electron experiences no force and
 exchanges no momentum,
energy, or angular momentum with the electromagnetic field; and in that the
 Hamiltonian, the
equations of motion, and the commutation relations involve no local
 contemporaneous Maxwell
field at the electron's position.

AB is a topological effect in that it requires the electron to be confined to a
 multiply-
connected region and in that there is no objective way to relate a phase shift
 to any
particular place
or to either arm of the interferometer.  The phase shift between any two
Feynman amplitudes
depends only upon the difference between the topological winding numbers $n$ of
 their paths.

\begin{eqnarray}
\delta \phi = (\delta n) {e\Phi \over \hbar c}\,.
\nonumber
\end{eqnarray}
The relative phase shift depends upon an integral whose integrand is not gauge
 invariant and not
observable.  The only gauge-invariant observable is the integral of this
 integrand over a closed
path, and its value is proportional to the magnetic flux enclosed by this path.
 This effect is
manifestly nonlocal, since its value depends upon a physical quantity in a
 region outside the
domain of integration.  It is topological in the sense that it depends only
upon the topology of the
path with reference to the enclosed magnetic flux.  In an interferometer, the
 winding numbers of
the two arms differ by unity.  The general role of the winding numbers is more
 obvious in the
magnetic scattering geometry, illustrated in Fig. 2.  The differences in phase
 shift between
different
paths are gauge invariant, but no measurable phase shift can be assigned to any
 one path because
${\int}\mbox{\boldmath $A$} \cdot d \mbox{\boldmath $r$}$
along any one path depends upon the choice of gauge.  The same
is true of the electric AB
effect.  The potential difference ${\Delta V}$ is gauge invariant, but the
 potential V on one
of the cylinders
can be given any value by choice of gauge.  Therefore there is no objective way
 to associate the
phase shift with one arm of the interferometer or the other.

\section{The Scalar Aharonov-Bohm effect with polarized neutrons}

In a recent series of experiments,
Allman $\it{et\,al.}$\cite{allman92,allman93}
passed unpolarized neutrons through a
Mach-Zehnder interferometer one arm of which traversed a magnetic field
${\bf  B}$, as shown
schematically in Fig. 3.  The intensities of the two outbound beams were
 observed to obey Eq.
(1.5), where now the relative phase shift is given by

\begin{equation}
\delta \phi = {\mu \over \hbar} B \tau \,,
\end{equation}
where $\mu$ is the neutron's magnetic moment, B  is the magnetic field
strength,
 and $\tau$
is the time
spent in the magnetic field.  The experimenters interpreted their results as
 demonstrating a new
topological effect which they named Scalar Aharonov-Bohm effect (SAB).  That
 claim was refuted
by one of us,\cite{peshkin92} who pointed to ambiguities introduced by the use
 of unpolarized
neutrons.  (The
same point had been made earlier by Zeilinger,\cite{zelinger86} and the meaning
 of this
kind of experiment was also discussed by Anandan.\cite{anandan82,anandan89})

Here we will analyze the ideal SAB experiment, also illustrated by Fig. 3, in
 which: the
neutron is to be polarized with ${\sigma_z}=+1$, where the $z$ direction is
that
 of the magnetic
field,
assumed to be spatially uniform; B(t) vanishes except during a time interval of
 length $\tau$
when it has
the value B; and the neutron is assumed to be in the magnetic field region
 throughout the time
interval t so that it never experiences a field gradient.  The relation of the
 $z$ direction to
the plane of
Fig. 3 is immaterial.  The purposes of this analysis are to show that using
 polarized neutrons
will not help and to explain how SAB differs in principle from AB.

	In SAB, the Hamiltonian

\begin{equation}
H = {{\mbox{\boldmath$p$}}^2 \over 2m} - \mu {\mbox{\boldmath$\sigma$}} \cdot
{\mbox{\boldmath$B$}}(t)
\end{equation}
contains the Maxwell field \mbox{\boldmath $B$}, in contrast to AB, where the
 Hamiltonian (1.1)
contains only the
gauge fields.  The operator equations of motion

\begin{equation}
{\hbar \over 2}\,\dot{\mbox{\boldmath$\sigma$}} = \mu{\mbox{\boldmath$\sigma$}}
 \times
{\mbox{\boldmath$B$}}(t)
\end{equation}
contain the local contemporaneous Maxwell field, in contrast to AB, where no
 electromagnetic
field enters the equation of motion of any measurable quantity.

However, it is argued that SAB resembles the electric AB affect (EAB) in that
 role of
\mbox{\boldmath $ B$} in the SAB Hamiltonian (2.2) is very much that of a
 potential acting
on the magnetic moment and in
that no force acts on the neutron.  Also, the consequences of Eq. (2.3) are
 possibly uncertain
because $\langle{\sigma_x}(t)\rangle=\langle{\sigma_y}(t)\rangle=0$ in a state
 with
${\sigma_z}=+1$.
In terms of the Schroedinger equation, one
may replace ${\sigma_z}$ by the number +1 in the Hamiltonian (2.2) so that it
 becomes

\begin{equation}
H = {{\mbox{\boldmath$p$}}^2 \over 2m} - \mu B(t)\,,
\end{equation}
and restrict the Hilbert space to what appears as a one-component wave function
 with no
dynamical
variables other that \mbox{\boldmath $x$} and \mbox{\boldmath $v$}.  Then the
 mathematical
analogy with EAB is complete and one has
the illusion \cite{allman92,allman93} that SAB is a nonlocal, topological
effect
 in the
same sense as is EAB.

That reasoning gets the correct phase shift but it leads to an incorrect
 interpretation of the
experiment.  In SAB, the relative phase shift depends upon an integral whose
 integrand is
locally
gauge invariant and observable at every point in the path of the neutron.  The
 integrand is
proportional to the magnetic field directly in the path of the neutron and does
 not depend
upon a
physical quantity in a region outside that path.  SAB does not have the same
 topological
character
as AB, because the SAB phase shift depends upon the local field along the path
 and not upon any
winding number expressing the topology of a path around a region in which the
 particle does not
move.  The operator equations of motion do involve the local, contemporaneous
 Maxwell field.

Moreover, in quantum mechanics, the spin is a dynamical variable and it cannot
 simply be
replaced by a number.  The right hand side of Eq. (2.3) is a torque
 \mbox{\boldmath $L$} on
the neutron whose expectation value vanishes at all times but whose
fluctuations
 do not vanish.

\begin{eqnarray}
\langle L_x \,\rangle &=& \langle L_y\,\rangle = 0
\nonumber \\
\langle L_{x}^{2}\, \rangle &=& \langle L_{y}^{2} \,\rangle = (mB)^2
\end{eqnarray}

Then an equal and opposite angular momentum must be transmitted to the local
electromagnetic field, again with zero expectation but with fluctuations
 correlated with those of the
neutron's angular momentum so that the total angular momentum is conserved.
 Those field
angular momentum fluctuations are not observable by a measurement on the field
 in the limit of a
classical field, but they are observable in principle in a finite field.

The effect of the torque on the neutron is exposed by considering the spin
 autocorrelation
operators

\begin{eqnarray}
C(t) &=& \case{1}{4} [{\sigma_x}\,(0) {\sigma_x}\,(t) + {\sigma_y}\,(0)
 {\sigma_y}\,(t) + h.c.]
\nonumber \\
S(t) &=& \case{1}{4} [{\sigma_x}\,(0) {\sigma_y}\,(t) - {\sigma_y}\,(0)
 {\sigma_x}\,(t) + h.c.]
\end{eqnarray}
These are Hermitean operators, measurable in principle, and they commute with
 so there is no
question about their significance in a state of definite ${\sigma_z}$.  Their
 equations of
motion,

\begin{eqnarray}
\dot{C} (t) &=& {2\mu B \over \hbar} S(t)
\nonumber \\
\dot{S} (t) &=& {2\mu B \over \hbar} C(t)\,,
\end{eqnarray}
contain the local contemporaneous Maxwell field and the solutions are given by

\begin{eqnarray}
	C(t) &=& cos(\omega t)
\nonumber \\
	S(t) &=& - sin(\omega t) \,,
\end{eqnarray}
where

\begin{equation}
\omega = 2\mu B/\hbar \,.
\end{equation}
These spin correlation operators cannot be described classically for spin 1/2,
 but they can be
described simply in the context of the usual semiclassical vector model.
There,
 the vectors
$\mbox{\boldmath $\sigma$} (0)$ and $\mbox{\boldmath $\sigma$} (t)$
are depicted as precessing on a cone with random phase so that  their
 projections on the xy
plane vanish on the average.  Equations (2.7, 2.8) show that the relative angle
$\vartheta (t) = \omega t$
between the two projections is changed by the action of the local torque.

When the two partial waves merge at the final mirror of the interferometer in
 Fig. 3, their
spin correlation angle is

\begin{equation}
\vartheta (\tau) = \omega \tau = 2 \delta \phi\,.
\end{equation}
The intensities in the two outgoing beams are of course given by the same Eqs.\
 (1.5).  However,
now the effect has been described as the measurement of a spin correlation.
The
 factor 2 in
Eq. (2.10) is the usual factor for rotations of spin 1/2.

None of this is really surprising from either a classical or a quantum
 mechanical point of
view.  A spinning particle is represented classically as a symmetric rotor
whose
 angular
momentum
precesses in a magnetic field.  The precession frequency $\omega$ is
independent
 of the angle
between the
rotation axis and the magnetic field.  That is why the spin autocorrelations
are
 independent
of the
spin state in Eqs. (2.8).  Classically, the only exceptions are the two states
 wherein
the spin points
exactly in the $+z$ or the $-z$ direction, a set of measure zero for which the
x
 and y
components vanish
and the precession frequency has no meaning.  However, if one defines the
 precession frequency
by any limiting process, it again has the value $\omega$.  In quantum mechanics
 only the
expectation
values of ${\sigma_x}$ and ${\sigma_y}$ vanish.  Their fluctuations are large,
equal in magnitude to ${\sigma_z}$.  In quantum
mechanics, the local magnetic field separates the energies of the two states of
 definite
${\sigma_z}$ and that
energy separation gives rise to the precession of\,\,${\sigma_x}$ and
 ${\sigma_y}$
which becomes visible in the spin autocorrelation functions.

\section{Conclusions}

	The Scalar Aharonov-Bohm effect has been described as the ordinary action of a
 magnetic
field on the magnetic moment of the neutron, causing the neutron to precess in
 the ordinary way.
The return torque transmits angular momentum to the local contemporaneous
 magnetic field in the
ordinary way.  Locality in the sense of Faraday and Maxwell is preserved to the
 extent that
it ever is in quantum mechanics.

We have identified measurable dynamical variables, the spin autocorrelation
 operators,
whose operator equations of motion obey the classical laws.  The conventional
 semiclassical vector
model shows exactly how the torque in the magnetic field acts on the spin
 autocorrelation.

SAB is not a topological effect in the same sense as is the AB effect, in spite
 of the
mathematical similarity of SAB and electric AB effect.  In SAB, we know exactly
 where the
neutron experienced the torque that changed the outcome of the experiment, and
 no gauge
transformation can obscure that information.

Allman $\it {et\,al.}$ \cite{allman93} defined a topological effect as one in
 which the
relative phase shift $\delta \phi$ is
independent of the energy of the neutron.  That criterion was justified by a
 result of
Zeilinger\cite{zelinger86,badurek93},
who however showed only that the energy independence is a necessary condition
 for a force-free
effect.

The trouble with using that criterion in the present context can be seen by
 considering a
problem in which the magnetic field in one arm of the interferometer is
replaced
 by an optical
phase
shifter whose index of refraction is made to depend upon the time and to differ
 from unity only
during the time the neutron is inside some box, for instance by pumping a
 refractive gas in and
out.  In principle, the phase shift can be made independent of the energy over
 the experimental
range.  No electromagnetic field is involved.  The energy-independence
criterion
 would
describe the influence of that phase shifter as a topological effect.

We have chosen to discuss the Aharonov-Bohm effect on the magnetic moment of a
 spin-
1/2 particle in terms of the SAB effect because of the experimental interest in
 that example.
However the discussion is identical for the Aharonov-Casher (AC) effect
 \cite{aharonov84}.
In AC, a neutron with ${\sigma_z} = +1$
traverses an external electric field in the $xy$ plane.  In an adequate
 approximation, the AC
Hamiltonian is given by Eq.J(2.2), where now $\bf B$ is the magnetic field in
 the rest frame
of the neutron, given by

\begin{equation}
{\mbox{\boldmath$B$}} = {{\mbox{\boldmath$p$}} \over mc} \times
{\mbox{\boldmath$E$}}({\mbox{\boldmath$r$}})\,.
\end{equation}
For a neutron whose velocity is confined to the $xy$ plane, \mbox{\boldmath
$B$}
 points in
the $z$ direction and interference effects not ascribable to forces, like
those in SAB, are predicted.  However, the torques, spin autocorrelations,
and angular momentum exchange with the local Maxwell field appear to be the
same as in SAB, so it follows that AC, like SAB, is neither a nonlocal nor a
topological effect.

The basic physics underlying our argument is in fact very simple.
The spin of a neutron precesses in an external magnetic field as a
result of the local interaction of the neutron magnetic moment with the
field. This precession has been observed in many experiments. It is
conjectured that such precession is absent when the neutron spin is
exactly in the directions of the field and the components of the spin
normal to the field vanish exactly; e.g. for a field in the
z-direction the spin components satisfy the condition
\begin{equation}
\sigma_x = \sigma_y= 0
\end{equation}
However this condition can be
satisfied in classical mechanics only for a set of states of measure
zero. In quantum mechanics this condition cannot be satisfied at all,
since the operators $\sigma_x$ and $\sigma_y$ do not commute with one
another and furthermore do not have an allowed zero eigenvalue.

The expectation values of $\sigma_x$ and $ \sigma_y$
do indeed vanish when a neutron is ``polarized in the z-direction";
i.e. when it is in an eigenstate of $\sigma_z$. However, this only means
that their average value vanishes. We have shown here
that the precession in the magnetic field of the spin components normal
to the field is still observable, even when the neutron is so-called
``polarized in the direction of the field". This precession is in fact
observed experimentally in the AC and SAB effects.
We have pointed out a marked
difference between the topology and locality which characterize the
AB and the analogous considerations in AC and SAB.
Instead the SAB experiment provides evidence that the normal components
of the neutron spin do indeed precess with the normal precession
frequency in an external magnetic field, even though the expectation
values of these normal components vanish. The precession is expressed
formally by spin autocorrelation functions.

This work is supported in part by the U. S. Department of Energy, Nuclear
Physics Division, under contract W-31-109-ENG-38.

\newpage
\begin{figure}
\caption{(a) Magnetic Aharonov-Bohm effect.  The shaded area is a
solenoid.  \newline (b) Electric Aharonov-Bohm effect.}
\end{figure}

\begin{figure}
\caption{Three Feynman paths from $X_1$ to $X_2$ with winding numbers.}
\end{figure}

\begin{figure}
\caption{Interferometer for polarized neutrons.  The shaded area is the
magnetic field region.}
\end{figure}


\newpage

\begin{references}
\bibitem{aharonov59} Y. Aharonov and D. Bohm, Phys. Rev. {\bf 115}, 485 (1959).
\bibitem{peshkin89} M. Peshkin and A. Tonomura, "The Aharonov-Bohm Effect",
       Lecture Notes in Physics No. {\bf 340} (Springer-Verlag 1989).
\bibitem{allman92} B. E. Allman {\it et al}., Phys. Rev. Lett. {\bf 68}, 2409
(1992).
\bibitem{allman93} B. E. Allman {\it et al}., Phys. Rev. A {\bf 48}, 1799
 (1993).
\bibitem{peshkin92} M. Peshkin, Phys. Rev. Lett. {\bf 69}, 2017 (1992).
\bibitem{zelinger86} A. Zeilinger in "Fundamental Aspects of Quantum Theory",
NATO ASI Series B, Vol. {\bf 144}, eds. V. Gorini and A. Frigerio (1986)
pp. 331.
\bibitem{anandan82} J. Anandan, Phys. Rev. Lett. {\bf 24}, 1660 (1982).
\bibitem{anandan89} J. Anandan in "Proceedings of the 3rd International
Symposium  Foundations of Quantum
Mechanics in the Light of New Technology", Physical Society of Japan, eds. S.
Kobayashi, H.
Ezawa, Y. Murayama, and S. Nomura (1989) pp. 98.
\bibitem{badurek93} G. Badurek et al., Phys. Rev. Lett. {\bf 71}, 307 (1993).
\bibitem{aharonov84} Y. Aharonov and A. Casher, Phys. Rev. Lett. {\bf 53},
319 (1984); see also A. S. Goldhaber, Phys. Rev. Lett. {\bf 62}, 482 (1989).
\end{references}
\end{document}